\documentclass[twocolumn]{openjournal}

\usepackage{amsmath}
\usepackage{hyperref}
\hypersetup{colorlinks=true, linkcolor=magenta, citecolor=magenta, filecolor=magenta, urlcolor=magenta}
\usepackage{orcidlink}
\graphicspath{{./}{./plots/}}

\begin{document}

\title{Direct Photometric Redshift Calibration of Weak Lensing Surveys for the DESI DR2 Analysis}

\author{A.~V.~Fisher\,\orcidlink{0009-0005-0530-0605},$^{1}$
J.~U.~Lange\,\orcidlink{0000-0002-2450-1366},$^{1}$
C.~Blake\,\orcidlink{0000-0002-5423-5919},$^{2}$
X.~Chen\,\orcidlink{0000-0003-3456-0957},$^{3,4,5,6}$
Z.~Chen,$^{7}$
J.~Choppin de Janvry\,\orcidlink{0009-0008-8066-446X},$^{8,9,10}$
S.~Heydenreich\,\orcidlink{0000-0002-7273-4076},$^{11}$
A.~Krolewski,$^{12,13,14}$
A.~Porredon\,\orcidlink{0000-0002-2762-2024},$^{15,16,17,6}$
R.~Ruggeri\,\orcidlink{0000-0002-0394-0896},$^{18}$
A.~Semenaite,$^{2}$
Z.~Shao\,\orcidlink{0000-0002-4585-3985},$^{19}$
A.~Souki\,\orcidlink{0009-0000-4771-7728},$^{15}$
Y.~Wang,$^{20}$
W.~Xu,$^{21}$
J.~Aguilar,$^{9}$
S.~Ahlen\,\orcidlink{0000-0001-6098-7247},$^{22}$
A.~Aviles\,\orcidlink{0000-0001-5998-3986},$^{23,24}$
F.~Beutler\,\orcidlink{0000-0003-0467-5438},$^{16}$
D.~Bianchi\,\orcidlink{0000-0001-9712-0006},$^{25,26}$
D.~Brooks,$^{27}$
A.~Carnero Rosell\,\orcidlink{0000-0003-3044-5150},$^{28,29}$
F.~J.~Castander\,\orcidlink{0000-0001-7316-4573},$^{30,31}$
E.~Chaussidon\,\orcidlink{0000-0001-8996-4874},$^{9}$
T.~Claybaugh,$^{9}$
A.~Cuceu\,\orcidlink{0000-0002-2169-0595},$^{9}$
A.~de la Macorra\,\orcidlink{0000-0002-1769-1640},$^{32}$
B.~Dey\,\orcidlink{0000-0002-5665-7912},$^{33,34}$
P.~Doel,$^{27}$
A.~Font-Ribera\,\orcidlink{0000-0002-3033-7312},$^{35,36}$
J.~E.~Forero-Romero\,\orcidlink{0000-0002-2890-3725},$^{37,38}$
E.~Gaztañaga\,\orcidlink{0000-0001-9632-0815},$^{30,39,31}$
S.~{Gontcho A Gontcho}\,\orcidlink{0000-0003-3142-233X},$^{40}$
G.~Gutierrez,$^{41}$
B.~Hadzhiyska\,\orcidlink{0000-0002-2312-3121},$^{42,43}$
C.~Hahn\,\orcidlink{0000-0003-1197-0902},$^{44}$
H.~K.~Herrera-Alcantar\,\orcidlink{0000-0002-9136-9609},$^{45,3,4,46,6}$
K.~Honscheid\,\orcidlink{0000-0002-6550-2023},$^{3,5,6}$
S.~Juneau\,\orcidlink{0000-0002-0000-2394},$^{47}$
T.~Karim\,\orcidlink{0000-0002-5652-8870},$^{33}$
D.~Kirkby\,\orcidlink{0000-0002-8828-5463},$^{48}$
A.~Kremin\,\orcidlink{0000-0001-6356-7424},$^{9}$
O.~Lahav\,\orcidlink{0000-0002-1134-9035},$^{27}$
C.~Lamman\,\orcidlink{0000-0002-6731-9329},$^{6}$
M.~Landriau\,\orcidlink{0000-0003-1838-8528},$^{9}$
L.~Le~Guillou\,\orcidlink{0000-0001-7178-8868},$^{49}$
A.~Leauthaud\,\orcidlink{0000-0002-3677-3617},$^{11,50}$
M.~E.~Levi\,\orcidlink{0000-0003-1887-1018},$^{9}$
M.~Manera\,\orcidlink{0000-0003-4962-8934},$^{51,36}$
A.~Meisner\,\orcidlink{0000-0002-1125-7384},$^{47}$
R.~Miquel\,\orcidlink{0000-0002-6610-4836},$^{35,36}$
J.~Moustakas\,\orcidlink{0000-0002-2733-4559},$^{52}$
A.~D.~Myers,$^{53}$
S.~Nadathur\,\orcidlink{0000-0001-9070-3102},$^{39}$
J.~ A.~Newman\,\orcidlink{0000-0001-8684-2222},$^{34}$
E.~Paillas\,\orcidlink{0000-0002-4637-2868},$^{54}$
W.~J.~Percival\,\orcidlink{0000-0002-0644-5727},$^{12,13,14}$
F.~Prada\,\orcidlink{0000-0001-7145-8674},$^{55}$
I.~P\'erez-R\`afols\,\orcidlink{0000-0001-6979-0125},$^{56}$
C.~Ravoux\,\orcidlink{0000-0002-3500-6635},$^{57}$
G.~Rossi,$^{58}$
L.~Samushia\,\orcidlink{0000-0002-1609-5687},$^{59,60}$
E.~Sanchez\,\orcidlink{0000-0002-9646-8198},$^{15}$
C.~Saulder\,\orcidlink{0000-0002-0408-5633},$^{61}$
D.~Schlegel,$^{9}$
M.~Schubnell,$^{62,63}$
M.~Siudek\,\orcidlink{0000-0002-2949-2155},$^{31,29}$
G.~Tarl\'{e}\,\orcidlink{0000-0003-1704-0781},$^{63}$
R.~Vaisakh\,\orcidlink{0009-0001-2732-8431},$^{64}$
and B.~A.~Weaver$^{47}$
}
\email{E-mail: af9886a@american.edu}
\shortauthors{A.~V. ~Fisher et al.}

\begin{abstract}
    Photometric redshift calibration is a critical step in dark energy studies using weak gravitational lensing. We present a unified photometric redshift calibration of the newest data releases of the four leading stage-III weak lensing surveys: the Dark Energy Camera All Data Everywhere (DECADE) survey, the Dark Energy Survey (DES), the Hyper Suprime-Cam (HSC) survey, and the Kilo-Degree Survey (KiDS). Our analysis builds on previous work by \citet{Lange2026_ApJ_998_263} which uses high-quality redshifts from the Dark Energy Spectroscopic Instrument (DESI) and neural network-based importance weights. In this work, we include DECADE for the first time, update the HSC and KiDS datasets to their latest public releases (HSC Y3 and KiDS-Legacy), expand the spectroscopic calibration sample by adding the zCOSMOS-bright catalog, and include estimates of uncertainties due to the choice of redshift sample. We obtain $n(z)$ constraints with statistical redshift uncertainties on the order of $\sigma_{\bar z} \approx 0.01$. This analysis approach provides a useful cross-check of existing calibration methods in light of recent cosmological tensions. We find modest improvements in spectroscopic coverage from the introduction of zCOSMOS, smaller magnitude shifts in redshift distributions relative to the shifts suggested by the HSC Y3 cosmic shear analysis, mild disagreement with the published KiDS-Legacy results, and overall consistency with the DECADE and DES Y3 analyses. Finally, we discuss the cosmological implications of our results, including their potential impact on measurements of $S_8$ and future dark energy constraints. This work provides photometric redshift calibrations for the joint analysis of DESI DR2 with weak lensing surveys.
\end{abstract}


\maketitle

\section{Introduction}

Weak gravitational lensing by large-scale structure is one of the primary probes of cosmology. It provides insight into the growth of structure in the Universe over time. Over the past two decades, increasingly precise measurements from wide-field imaging surveys have constrained measurements of dark energy and tested the standard cosmological model. Current stage-III surveys, including the Dark Energy Camera All Data Everywhere survey \citep[DECADE; ][]{Anbajagane2025_OJAp_8_6159}, the Dark Energy Survey \citep[DES; ][]{Flaugher2015_AJ_150_150, Myles2021_MNRAS_505_4249, Abbott2022_PhRvD_105_3520}, the Hyper Suprime-Cam survey \citep[HSC; ][]{Sugiyama2023_PhRvD_108_3521, Rau2023_MNRAS_524_5109, Dalal2023_PhRvD_108_3519, Aihara2018_PASJ_70_4}, and the Kilo-Degree Survey \citep[KiDS; ][]{Kuijken2015_MNRAS_454_3500, Asgari2021_AA_645_104, Wright2025_AA_703_144, Wright2025_AA_703_158}, have measured galaxy weak lensing across large areas of sky, while upcoming stage-IV surveys are predicted to provide an unprecedented increase in statistical precision \citep{TheLSSTDarkEnergyScienceCollaboration2018_arXiv_1809_1669, Akeson2019_arXiv_1902_5569, EuclidCollaboration2025_AA_697_1}. As statistical uncertainties of weak lensing measurements continue to decrease, it is increasingly critical to put stringent requirements on potential systematic errors.

Weak lensing distortions are sensitive to the intervening mass distribution as well as the distance to the galaxy being lensed, i.e., its redshift. Spectroscopic redshifts provide precise measurements, but they are observationally expensive and infeasible for the vast number of faint galaxies used in weak lensing studies due to their bias towards bright, nearby galaxies. Surveys instead rely on photometric redshifts, which estimate galaxy distances from broad-band imaging data and allow for the targeting of many more galaxies to measure weak gravitational lensing. However, photometric redshifts are subject to bias that can propagate into cosmological parameter inferences. This means weak lensing analysis requires robust calibration of the redshift distribution $n(z)$ for galaxies in each tomographic bin \citep{Huterer2006_MNRAS_366_101, Hildebrandt2012_MNRAS_421_2355, Benjamin2013_MNRAS_431_1547, Huterer2013_MNRAS_432_2945, Samuroff2017_MNRAS_465_20, Joudaki2020_AA_638_1, Tessore2020_OJAp_3_6}. Tomographic bins are groupings of galaxies, most of which fall within a chosen redshift range, and are typically characterized through the mean redshift $\bar{z}$ with more distant galaxies in higher bins \citep{Zhang2023_MNRAS_518_709}.

Several complementary approaches have been developed for photometric redshift calibration \citep[see][and references therein]{Newman2022_ARAA_60_363}. These methods include direct calibration techniques, which rely on mapping a photometric galaxy population to a known spectroscopic sample, and clustering redshift techniques that infer redshift distributions from angular correlations with tracers of known redshift distributions. Direct calibration methods use spectroscopic and often high-fidelity photometric redshift samples to estimate the $n(z)$ of a source population. Spectroscopic samples are typically incomplete and subject to complex selection effects. Approaches for mitigating these biases typically include self-organizing map \citep[SOM;][]{Kohonen2001_som_book} methods as a commonly used tool \citep{Masters2019_ApJ_877_81, Buchs2019_MNRAS_489_820, Wright2020_AA_637_100}. Another approach is clustering redshift calibration, that uses the fact that galaxies at similar redshifts reside in the same underlying large-scale structure. One can measure the angular cross-correlation between a photometric sample of unknown redshift and a reference sample with known redshifts to reconstruct the $n(z)$ of a photometric sample \citep[see, e.g.,][]{Newman2008_ApJ_684_88, Gatti2022_MNRAS_510_1223, dAssignies2025_AA_702_155, ChoppindeJanvry2026_JCAP_05_073, Ruggeri2026_OJAp_9_3639}. Contrary to direct calibration, clustering methods require accurate modeling of galaxy bias evolution and are sensitive to observational systematics affecting clustering estimates.

As stated, one difficulty of direct photometric redshift calibration is that spectroscopic samples are not fully representative of the selected weak lensing galaxy population. Spectroscopic surveys preferentially target bright, nearby galaxies and those with strong spectral features, leading to an incomplete and biased coverage, especially of faint sources which contribute significantly to weak lensing measurements. This incompleteness can introduce biases in the inferred redshift distributions, and ultimately cosmology, if not properly mitigated. Recently, \citet{Lange2026_ApJ_998_263} introduced a unified calibration framework based on importance weights derived from neural networks. This has enabled consistent estimation of $n(z)$ across multiple surveys while, for example, not being subject to the pixelation effects inherent to SOM-based methods \citep{Wright2020_AA_637_100, Wright2025_AA_703_144}. The pipeline uses high-fidelity samples from the Dark Energy Spectroscopic Instrument \citep[DESI; ][]{Ratajczak2026_AJ_171_71} to construct representative calibration catalogs. Such independent redshift calibrations are especially important in light of current cosmological tensions, such as discrepancies in measurements of $S_8$ \citep[see][and references therein]{Abdalla2022_JHEAp_34_49}, where systematic shifts in redshift calibration at the level of $\Delta \bar{z} \sim 0.01$ can have non-negligible impact on inferred cosmological parameters.

DESI \citep{Levi2013_arXiv_1308_0847, DESICollaboration2022_AJ_164_207, DESICollaboration2016_arXiv_1611_0036, Schlafly2023_AJ_166_259, Guy2023_AJ_165_144, Miller2024_AJ_168_95, Poppett2024_AJ_168_245, DESICollaboration2026_AJ_171_285} has recently produced its first cosmological results based on Data Release 1 (DR1) observations and is now working on the cosmology analysis of DESI Data Release 2 (DR2) data products \citep[see, e.g.,][]{AbdulKarim2025_PhRvD_112_3515}. For the DESI DR2 galaxy lensing analysis, DESI will combine spectroscopic galaxy clustering, weak gravitational lensing, and cosmic shear in a so-called $3\times2$ point framework to place improved constraints on cosmology. We produce this work's redshift calibration by expanding on the methods of \citet{Lange2026_ApJ_998_263}. Compared to \citet{Lange2026_ApJ_998_263}, we added the calibration of the DECADE lensing catalog, and replace HSC Year 1 (Y1) and KiDS-1000 with the HSC Year 3 (Y3) and KiDS-Legacy data sets, respectively. In addition to the spectroscopic data from DESI used in the previous work, we incorporate the zCOSMOS-bright spectroscopic sample \citep{Lilly2007_ApJS_172_70, Lilly2009_ApJS_184_218}, a high-fidelity redshift catalog in the COSMOS field, into the tiered calibration pipeline. The idea behind the inclusion of zCOSMOS is an improvement in coverage of the galaxy population in regions of color space that are sparsely sampled by DESI alone, and a cross-check on potential systematic biases associated with reliance on a single spectroscopic training sample. Finally, we estimate the impact of the use of high-quality multi-band photometric redshifts for galaxies not probed by spectroscopic redshifts.

This paper is organized as follows. In Section~\ref{sec:data}, we describe the photometric and spectroscopic data sets used in this analysis, including DECADE, DES Y3, HSC Y3, KiDS-Legacy, DESI, and zCOSMOS. In Section~\ref{sec:methods}, we outline the calibration methodology and its implementation within the unified framework. Section~\ref{sec:verification} presents validation tests of the pipeline and Section~\ref{sec:results} contains our main results for the $n(z)$. We discuss the implications of our findings for weak lensing cosmology in Section~\ref{sec:discussion} and conclude in Section~\ref{sec:conclusion}.

\section{Data}
\label{sec:data}

In this section, we describe the data sets used. We begin with the construction of the redshift calibration catalog in the COSMOS field, followed by a description of the photometric lensing data from the four weak lensing surveys, DECADE, DES Y3, HSC Y3, and KiDS-Legacy. As detailed in \citet{Lange2026_ApJ_998_263}, the COSMOS field acts as the common calibration region due to the availability of multi-wavelength photometry and spectroscopic coverage \citep{Scoville2007_ApJS_172_1}.

\subsection{Calibration Data}

Our calibration catalog is constructed by combining multiple photometric and spectroscopic data sets in the COSMOS field. As a baseline, we use the FARMER COSMOS2020 photometric catalog \citep{Weaver2022_ApJS_258_11}, which provides high-quality multi-band photometry and photometric redshifts for a representative sample of galaxies. These redshifts are derived using the \textsc{LePhare} $\chi^2$ method \citep{Arnouts2011_ascl_soft_8009}, producing \texttt{lp\_zBEST}, which are photometric redshifts fitted to galaxy templates based on \citet{Polletta2007_ApJ_663_81, Ilbert2009_ApJ_690_1236, Ilbert2013_AA_556_55}. This catalog is supplemented with photometric measurements from the DESI Legacy Imaging Surveys DR10 \citep{Dey2019_AJ_157_168}, which provide uniform optical and infrared photometry across the DESI footprint, and serve as the primary feature space for our neural network model.

Spectroscopic redshift information comes from two complementary sources. First, we utilize the DESI-COSMOS spectroscopic catalog in the COSMOS region, which provides high-precision redshifts over a wide range of galaxy types. This public catalog contains $233,247$ unique spectra over a $8 \, \mathrm{deg}^2$ field \citep{Ratajczak2026_AJ_171_71}, and matches to the DESI Legacy Survey DR10 \citep{Dey2019_AJ_157_168}, HSC PDR3 \citep{Aihara2022_PASJ_74_247}, and COSMOS2020 photometry \citep{Weaver2022_ApJS_258_11}. We retain DESI targets with valid observations, identified by \texttt{QUALITY\_Z}, and when multiple DESI targets are associated with a COSMOS2020 source, we prioritize successful redshifts from DESI large-scale structure (LSS) target classes identified as the BGS, LRG, ELG, and QSO targeting flags, over other successful DESI redshifts. Second, we incorporate the zCOSMOS-bright spectroscopic survey \citep{Lilly2007_ApJS_172_70, Lilly2009_ApJS_184_218}, a high-fidelity redshift sample that is expected to improve completeness in regions of color space that are underrepresented in DESI. We select sources with \texttt{FlagS==1} and retain redshifts in confidence classes 3 or 4, which satisfies the high-confidence zCOSMOS quality criteria.

We match all photometric and spectroscopic catalogs using a $1 \, \mathrm{arcsec}$ radius and construct a unified catalog where each object is assigned a single best-available redshift following a hierarchical scheme:
\begin{equation}
    z_{\mathrm{best}} = 
    \begin{cases}
        z_{\mathrm{DESI}}, & \text{if available}, \\
        z_{\mathrm{zCOSMOS}}, & \text{else if available}, \\
        z_{\mathrm{COSMOS2020}}, & \text{otherwise}.
    \end{cases}
\end{equation}
The calibration catalog combines the statistical power of DESI with the high completeness of zCOSMOS and the depth of COSMOS2020 photometric redshifts.

\subsection{Dark Energy Camera All Data Everywhere}

We utilize the newest Dark Energy Camera All Data Everywhere shear catalog which covers $\sim 5,500 \, \mathrm{deg}^2$ of the Northern Galactic cap, imaging $107$ million galaxies. The construction and calibration of the weak lensing catalogs is very similar to that of the Dark Energy Survey. Source galaxies are grouped into four tomographic bins of width $\Delta z \sim 0.3$ via the use of SOM-based photometric redshifts. Wide-field photometry is constructed from DECam imaging in the $r i z$ bands. Redshift distributions are calculated through a combination of SOM-based redshift calibration, clustering redshifts, and synthetic source injection \citep{Anbajagane2025_OJAp_8_6159}. Unlike the DES analysis, the DECADE calibration does not explicitly correct for the impact of blending on the redshift distribution \citep{MacCrann2022_MNRAS_509_3371}.

\subsection{Dark Energy Survey}

We use the Year 3 data release of the Dark Energy Survey (DES Y3), which covers approximately $4,000 \, \mathrm{deg}^2$ of sky in the Southern Galactic Cap and provides a well-characterized weak lensing sample \citep{Gatti2021_MNRAS_504_4312}. Galaxies are assigned to four tomographic bins based on multi-band photometry in the $r i z$ bands \citep{Myles2021_MNRAS_505_4249}, and the survey includes detailed injection simulations \citep{Everett2022_ApJS_258_15} that enable estimation of selection effects. As in \citet{Lange2026_ApJ_998_263}, these simulations are used to determine the probability that galaxies in the COSMOS field would be included in the DES weak lensing sample and their corresponding statistical weights. The fiducial calibration utilizes a combination of SOM-based direct and clustering redshift calibration \citep{Myles2021_MNRAS_505_4249} and also accounts for the effects of blending \citep{MacCrann2022_MNRAS_509_3371}.

\subsection{Hyper Suprime-Cam}

We use the Hyper Suprime-Cam survey Year 3 data release (HSC Y3), which significantly expands the area of the sample relative to previous releases, at approximately $400 \, \mathrm{deg}^2$ of the Northern sky \citep{Li2022_PASJ_74_421}. HSC Y3 provides deep $grizy$ photometry with high effective galaxy number density, enabling precise weak lensing measurements \citep{Rau2023_MNRAS_524_5109, Dalal2023_PhRvD_108_3519}. The COSMOS field has been observed at a depth comparable to the full survey, allowing for direct mapping between COSMOS galaxies and the HSC weak lensing sample. HSC Y3 utilizes four equidistant tomographic bins spanning from redshift $0.3$ to $1.5$. Photometric redshift distributions are derived by stacking the photo-z posterior distributions and further calibrated with clustering redshifts via the CAMIRA LRG catalog to infer ensemble redshift distribution for bins 1 and 2 and partially for bin 3. However, the CAMIRA LRG catalog does not extend past $z > 1.2$, leading to large uncertainties in the mean redshifts of bins 3 and 4 and motivating our investigation of $S_8$ value differences between this method and the fiducial one in Sec.~\ref{sec:discussion}.

\subsection{Kilo-Degree Survey}

We use the KiDS-Legacy data set, which covers almost $1,400 \, \mathrm{deg}^2$ of sky in the Southern Hemisphere \citep{Wright2025_AA_703_158}. KiDS-Legacy provides multi-band photometry in the $ugriZYJHK_s$ bands extending from the optical to the near-infrared, enabling robust photometric redshift estimation. This analysis employs six equipopulated tomographic bins, as opposed to KiDS-1000 where five were used. Also different from the KiDS-1000 analysis, the KiDS-Legacy data release is sampled within the COSMOS field at survey depth, thus requiring no artificial methods to estimate lensing cuts and weights from raw photometric catalogs as in \citet{Lange2026_ApJ_998_263}. Fiducial redshift distributions are calculated via SOMs and simulations such as MICE2 and SKiLLS \citep{Carretero2015_MNRAS_447_646, Wright2025_AA_703_144, Li2023_AA_670_100}.

\section{Methods}
\label{sec:methods}

In this section, we briefly summarize the photometric redshift calibration methodology. Our approach closely follows the unified framework introduced in \citet{Lange2026_ApJ_998_263}, and we refer the reader there for a detailed description. Here, we highlight the key components of the pipeline and the modifications associated with the updated data sets and the inclusion of additional spectroscopic information. The central idea of the method is to estimate the $n(z)$ for each weak lensing sample by combining high-quality redshifts in the COSMOS field with a weighting scheme that accounts for both spectroscopic incompleteness and weak lensing selection effects. This is achieved by assigning each galaxy two weights: a completeness weight and a lensing weight.

\subsection{Calibration Framework}

We construct a unified calibration catalog in the COSMOS field by combining photometric and spectroscopic data as described in Section~\ref{sec:data}. Each galaxy is assigned a best-available redshift $z_{\mathrm{best}}$, prioritizing spectroscopic redshifts from DESI and zCOSMOS, followed by COSMOS2020 photometric redshifts. The primary modification we make is the inclusion of the zCOSMOS-bright spectroscopic sample as an additional tier in the calibration pipeline for reasons outlined in Section~\ref{sec:data}.

\subsection{Incompleteness Correction}

The spectroscopic calibration sample is not fully representative of the underlying galaxy population due to selection effects in fiber targeting, i.e., which galaxies were observed by DESI or zCOSMOS. To account for this, we estimate the probability $p_{\mathrm{f}}$ that a given galaxy is assigned a fiber. We model $p_{\mathrm{f}}$ using a neural network trained on photometric observables with hyper-parameters very similar to those in \citet{Lange2026_ApJ_998_263}. We estimate $p_{\mathrm{f}}$ using the \texttt{MLPClassifier} in \textsc{scikit-learn}. Input features include RA and DEC along with Legacy Survey photometry in $g,r,i,z$ and WISE $W_1, W_2$ bands. The target classification $F$ is whether any galaxy was targeted by DESI or zCOSMOS. Note, however, that this classification does not require that the redshift obtained was secure.

The resulting targeting probability is used to define a completeness weight $w_{\mathrm{c}}$ that upweights galaxies underrepresented in the spectroscopic sample. We adopt the same importance-weighting regime as in \citet{Lange2026_ApJ_998_263}, including the use of a minimum probability threshold to control statistical noise and mitigate the impact of poorly constrained regions of parameter space. This procedure is described in the following equation:

\begin{equation}
    w_\mathrm{c} =
    \begin{cases}
        p_\mathrm{f}^{-1} &\mathrm{ if } \, p_\mathrm{f} > p_\mathrm{f, min} \, \, \mathrm{ and } \, F \\
        0 &\mathrm{ if } \, p_\mathrm{f} > p_\mathrm{f, min} \, \, \mathrm{ and \, not } \, F \\
        1 &\mathrm{ if } \, p_\mathrm{f} \leq p_\mathrm{f, min} \\
    \end{cases}
    \label{eq:weights}
\end{equation}

Galaxies with targeting probabilities above $p_\mathrm{f,min}$ contribute to the calibration only if they were targeted spectroscopically by DESI or zCOSMOS. These galaxies are assigned a weight of $p_\mathrm{f}^{-1}$, which corrects for them representing a larger population of similar galaxies that were not targeted. Note that some of the galaxies in this group may have been targeted by DESI or zCOSMOS but not have obtained secure redshifts. In those cases, following our tiered redshift system, we use COSMOS2020 redshifts. Galaxies below the $p_\mathrm{f,min}$ threshold are always used and given a weight of unity, regardless of spectroscopic coverage, and they only rely on the best available redshift. $p_\mathrm{f,min}$ balances maximizing the fraction of galaxies calibrated with spectroscopic redshifts and maintaining sufficient effective sample size. Smaller values increase the spectroscopic contribution, but also the shot noise, while larger values rely more heavily on photometric redshifts, but improve statistical precision.

\subsection{Lensing Weights}

The above completeness weights account for spectroscopic incompleteness in the calibration sample, independent of the weak lensing surveys. So, we assign each galaxy a lensing weight that reflects the probability it enters a given sample and its contribution to the shear measurement in order to estimate the effective redshift distribution. For each weak lensing sample $b$, defined by a survey and tomographic bin, every COSMOS galaxy is assigned a lensing weight. This weight accounts for the fact that not every galaxy satisfies quality cuts or is assigned to the bin associated with $b$. This is described by the selection probability $w_{\mathrm{t}}(b)=p(b)$ and serves the same purpose as the so-called transfer function \citep{Myles2021_MNRAS_505_4249}.  In weak lensing surveys, galaxies are typically assigned an explicit weight $w_{\mathrm{s}}$, designed to reduce the measured shear variance. Finally, the measured ellipticity of a galaxy may have a non-unity response to intrinsic ellipticity changes caused by gravitational shear, i.e., $e_{\mathrm{obs}} = R \times e_{\mathrm{int}}$, where $R \neq 1$. This adds $w_{\mathrm{R}} = R$ as an additional weight to calculate the effective $n(z)$.

Combined with the factors listed above, we assign a lensing weight $w_{\mathrm{l}}(b)$ that captures both the probability a galaxy in COSMOS is included in the sample and its statistical contributions to shear measurements. This is defined as

\begin{equation}
    w_{\mathrm{l}}(b)=w_{\mathrm{t}}(b)\langle w_{\mathrm{s}}w_{\mathrm{R}}\rangle_b,
\end{equation}

where the expectation value $\langle\rangle_b$ is taken over all realizations where a galaxy could randomly be observed in weak lensing sample $b$. The estimation of lensing weights follows \citet{Lange2026_ApJ_998_263}, updated with the latest data releases.

\subsubsection{Dark Energy Survey Year 3 \& Dark Energy Camera All Data Everywhere Survey}

For DECADE and DES Y3, we use Balrog injection simulations \citep{Everett2022_ApJS_258_15} to estimate the selection probability $w_{\mathrm{t}}(b)$ and effective shear weight and response $\langle w_{\mathrm{s}}w_{\mathrm{R}}\rangle_b$ via Monte Carlo sampling. For DES Y3, the shear response, $R$, is taken using the trace of the response matrix \citep{Huff2017_arXiv_1702_2600}, i.e.,
\begin{equation}
    w_{\mathrm{R}} = 0.5(R_{11} + R_{22}).
\end{equation}
For DECADE, we follow the weighting scheme described in \citet{Anbajagane2025_OJAp_8_6159}. Similar to DES Y3, galaxies are weighted by the product of the shear weight $w_{\mathrm{s}}$ and the shear response $R$ from the \textsc{metacalibration} estimator \citep{Sheldon2017_ApJ_841_24, Huff2017_arXiv_1702_2600}. However, rather than using per-object values, $w_{\mathrm{s}}$ and $\mathrm{R}$ are averages computed on a grid of signal-to-noise and galaxy size relative to the point-spread function (PSF), following the procedure in \citet{Anbajagane2025_OJAp_8_6158}.

\subsubsection{Hyper Suprime-Cam Year 3}

For HSC Y3, COSMOS galaxies are observed directly at survey depth, so each galaxy has a single realization in the lensing catalog. In this case, the selection probability is binary, with $w_{\mathrm{t}}(b) \in \{0,1\}$ depending on whether the galaxy passes the lensing cuts and is assigned to the specific tomographic bin $b$. The shear weighting scheme is as follows
\begin{equation}
    w_{\mathrm{R}} = (1 - e_{\mathrm{rms}}^2)(1 + m),
\end{equation}
where $e_{\mathrm{rms}}$ is the root-mean-square intrinsic ellipticity per component and $m$ is the multiplicative shear bias. The factor $(1 - e_{\mathrm{rms}}^2)$ accounts for the shear response and $(1 + m)$ corrects for multiplicative calibration biases in the shear measurement \citep[see Sec.~2.4.2 of][]{Lange2024_OJAp_7_57}.

\subsubsection{Kilo-Degree Survey Legacy}

For KiDS-Legacy, COSMOS galaxies are also observed directly, so again, each galaxy has a single realization in the lensing catalog. The selection probability is therefore binary, as described above. Finally, KiDS-Legacy assumes the same response for all galaxies within a tomographic bin, implying $w_{\mathrm{R}} \equiv 1$.

\subsection{Redshift Distribution Estimation}

The redshift distribution for a given weak lensing sample $b$ is estimated by connecting the spectroscopic calibration sample (DESI + zCOSMOS) and the COSMOS2020 photometric sample to the weak lensing catalogs through their overlap in the COSMOS field.

We spatially match galaxies in the weak lensing catalogs to the COSMOS calibration catalog using a $1$ arcsecond matching radius and positions based on the COSMOS2020 sample. Only galaxies present in both catalogs are used in the calibration. Because the COSMOS photometric catalog is significantly deeper than the lensing surveys, the matching should not introduce lensing selection biases.

Each galaxy is assigned a total weight
\begin{equation}
    w_{\mathrm{tot}} = w_{\mathrm{c}} \, w_{\mathrm{l}}(b),
\end{equation}
which combines the completeness correction from the calibration sample with the correction and shear weighting of the weak lensing sample.

\subsection{Statistical Uncertainties}

We estimate statistical uncertainties following the formalism of \citet{Lange2026_ApJ_998_263}, accounting for both shot noise and cosmic variance in the COSMOS field. Shot noise is due to the COSMOS calibration field containing a finite number of galaxies and cosmic variance reflects fluctuations in the large-scale structure of the calibration field that causes it to differ from the cosmic mean. We approximate these as uncorrelated in order to treat them independently, so the uncertainty in the mean redshift of a weak lensing sample is expected to be
\begin{equation}
    \sigma_{\bar z}^2 = \sigma_{\bar z, \mathrm{SN}}^2 + \sigma_{\bar z, \mathrm{CV}}^2.
\end{equation}
The shot noise depends on the effective number of galaxies contributing to the weighted redshift distribution. Galaxies are assigned non-uniform lensing weights, so the statistical precision is determined from the effective sample size. This is estimated as
\begin{equation}
    \sigma_{\bar z, \mathrm{SN}}^2 \approx \frac{\sum w_i^2 (\bar z - z_i)^2}{(\sum w_i)^2} = \frac{\sigma_z^2}{N_\mathrm{eff}},
    \label{eq:shot_noise}
\end{equation}
where $\sigma_z$ is the standard deviation of the weighted redshift distribution and $N_{\mathrm{eff}}$ is the effective sample size described by
\begin{equation}
    N_{\mathrm{eff}} = \frac{(\sum w_i)^2}{\sum w^2_i}\frac{\sum w_i(z_i-\bar{z})^2}{\sum w_i}\frac{\sum w_i^2}{\sum w^2_i(z_i-\bar{z})^2}.
\end{equation}

The cosmic variance accounts for large-scale matter density fluctuations within the COSMOS calibration field. The covariance of matter overdensities between different redshifts \citep{Moster2011_ApJ_731_113} is calculated from the linear matter power spectrum with the weighted redshift distribution to estimate the uncertainty on $\bar{z}$ as described in \citet{Lange2026_ApJ_998_263}. Note that the cosmic variance estimate will depend on the bias $b_\mathrm{gal}$ of source galaxies.

The increased spectroscopic completeness from the addition of zCOSMOS leads to a modest increase in the fraction of $n(z)$ weight coming from spectroscopic sources and therefore a theoretical reduction in shot noise. Cosmic variance remains unchanged, as it is determined by the volume of the calibration field.

\subsection{Systematic Uncertainties}

In addition to statistical uncertainties, our analysis is also affected by systematic uncertainties related to the inevitable use of multi-band photometric redshifts from COSMOS2020. This may shift derived $n(z)$ in two ways. First, COSMOS2020 redshifts scatter around the true redshifts, leading to statistical noise. Second, and more importantly, photometric redshifts may be systematically biased. Unfortunately, to determine such a bias, one needs to analyze the difference between COSMOS2020 redshifts and spectroscopic redshifts over a color-representative sample. However, the lack of such a representative sample is exactly why we need to use COSMOS2020 in the first place. Ultimately, this suggests that estimates of the COSMOS2020 redshift bias are highly uncertain and rely on extrapolating from incomplete or biased data.

One approach taken in \cite{Yin2025_arXiv_2510_3566} is to sample the difference between COSMOS2020 redshifts and spectroscopic redshifts and assume the difference depends on COSMOS2020 redshift and magnitude. However, as just discussed, the sample of galaxies with both COSMOS2020 and spectroscopic redshifts is almost certainly biased in magnitude and color, probably leading to a biased estimate of the offset. Furthermore, there is only a limited number of samples, leading to increased uncertainties in the estimate of the bias. Instead, in this work, we use COSMOS2020 redshifts derived from \textsc{EAZY} \citep{Brammer2008_ApJ_686_1503}. As shown in \cite{Weaver2022_ApJS_258_11}, the scatter and bias between \textsc{EAZY} and \textsc{LePhare} redshifts is qualitatively and quantitatively similar to the difference between \textsc{LePhare} and spectroscopic redshifts. Furthermore, unlike spectroscopic samples, such a sample is color and magnitude complete.

To estimate the noise and bias due to the use of COSMOS2020, we measure the mean redshift difference between \textsc{EAZY} and \textsc{LePhare} redshifts and its corresponding uncertainty, appropriately weighted for each survey and tomographic bin. We add both numbers in quadrature since the first is the estimate of the bias and the second is an estimate of the scatter. Finally, this number multiplied by the fractional contribution of COSMOS2020 to the $n(z)$ serves our estimate for the redshift sample error $\sigma_{\bar z, \mathrm{RS}}$.

We briefly discuss the implications of the redshift sample error in Sec.~\ref{sec:discussion}.

\section{Verification}
\label{sec:verification}

The addition of zCOSMOS modifies the composition of the calibration catalog and changes the fraction of galaxies calibrated spectroscopically. Therefore, before determining effective redshift distributions for the different weak lensing samples, we verify the updated weighting scheme remains stable and accurately recovers unbiased $n(z)$. We perform four complementary validation tests:

\begin{itemize}
    \item We examine the impact of the fiber fraction $p_\mathrm{f}$ threshold on the spectroscopic completeness and statistical precision of the calibration.
    \item We verify the importance weighting procedure accurately reconstructs the underlying redshift distribution using COSMOS2020 photometric redshift distribution.
    \item We compare the redshift measurements of the different calibration catalogs to assess their mutual consistency.
    \item We assess the accuracy of our neural network by testing a gradient boosting regime to compare the $n(z)$ reconstruction abilities.
\end{itemize}

\subsection{Impact of Fiber Fraction Threshold}

\begin{figure*}
    \centering
    \includegraphics{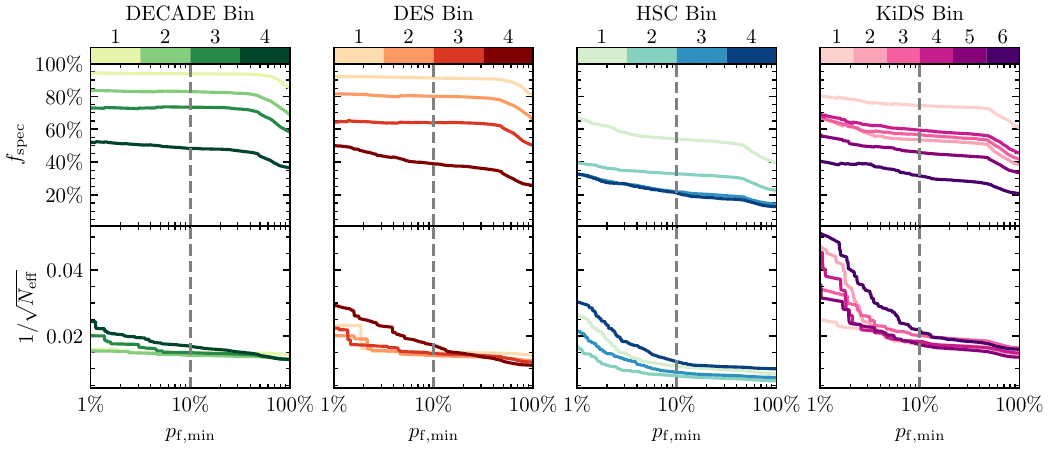}
    \caption{Dependence of the spectroscopic calibration fraction and shot-noise uncertainty on the minimum fiber-assignment probability, $p_\mathrm{f, min}$, for the four weak lensing samples. Columns show the results for DECADE, DES, HSC, and KiDS, respectively. In the upper panels, we plot the weighted fraction of galaxies in each calibration sample for which DESI obtained a successful redshift. Objects targeted by DESI or zCOSMOS that did not obtain a successful redshift are treated as not having a spectroscopic redshift and therefore reduce $f_\mathrm{spec}$. The lower panels show the inverse square root of the effective sample size, which characterizes the shot-noise uncertainty. The gray vertical line marks our fiducial choice of $p_\mathrm{f, min}=0.10$.}
    \label{fig:f_desi}
\end{figure*}

The fiber fraction threshold $p_\mathrm{f, min}$ determines how strongly the redshift calibration relies on redshifts from DESI and zCOSMOS as opposed to high-quality multi-band photometric redshifts from COSMOS2020. Galaxies with targeting probabilities above the threshold contribute only if they were spectroscopically observed and are upweighted to account for incompleteness, while galaxies below the threshold are used regardless of spectroscopic coverage. Decreasing $p_\mathrm{f, min}$ increases the fraction of spectroscopic reliance at the cost of increasing statistical noise due to larger importance weights.

In Fig.~\ref{fig:f_desi}, we observe how variations in $p_\mathrm{f, min}$ affect the relative contribution of spectroscopic redshifts and shot noise for DECADE, DES, HSC, and KiDS. Uppermost panels show the weighted average of galaxies with DESI redshifts where the weight is $w_\mathrm{tot}$. Galaxies observed by DESI that did not obtain a spectroscopic redshift will not be counted in this fraction. For all weak lensing samples, reducing $p_\mathrm{f, min}$ increases the fraction of galaxies with DESI redshifts as expected. It can also be seen that shallower DECADE, DES, and KiDS surveys can be calibrated with a much higher fraction of DESI galaxies than the deeper HSC survey. Higher-redshift tomographic bins have fewer contributions from galaxies with DESI redshifts. Lower panels show the shot noise as measured by the effective sample size. It can be seen that lowering $p_\mathrm{f, min}$, signifying using fewer galaxies with only high-quality photometric redshifts, leads to an increase in shot noise.

Compared to \citet{Lange2026_ApJ_998_263}, the dependence on $p_\mathrm{f, min}$ is noticeably weaker across all four surveys. This behavior is expected because zCOSMOS increases the spectroscopic completeness of the calibration catalog over a broad range of galaxy properties. Changing $p_\mathrm{f, min}$ has a smaller effect on the fraction of galaxies calibrated spectroscopically and the effective sample size than in the DESI-only calibration.

\subsection{Impact of Weighting}
\label{subsec:weighting}

\begin{figure*}[h!]
    \centering
    \includegraphics{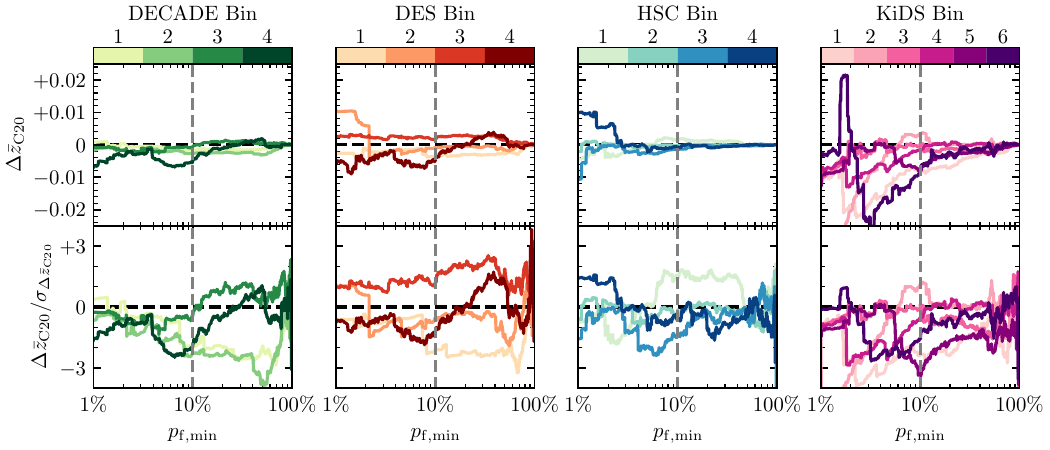}
    \caption{Test of the importance-weighting method using COSMOS2020 photometric redshifts. The plotted quantity is the change in the weighted mean COSMOS2020 redshift relative to $p_\mathrm{f,min}=100\%$, shown as a function of $p_\mathrm{f,min}$. Following \citet{Lange2026_ApJ_998_263}, the $p_\mathrm{f,min}=100\%$ case represents an unweighted sample containing all galaxies in the COSMOS2020 catalog. The four columns correspond, from left to right, to DECADE, DES, HSC, and KiDS. The upper panels display the absolute redshift difference, while the lower panels give the corresponding significance, defined as the difference divided by its uncertainty after accounting for shot noise.}
    \label{fig:d_z}
\end{figure*}

The goal of the calibration pipeline is an unbiased recovery of $n(z)$ for each weak lensing sample despite spectroscopic incompleteness. To test this, we perform an internal consistency check using the COSMOS2020 photometric redshifts. Nearly all galaxies in the COSMOS field have high-quality COSMOS2020 redshifts, so the full catalog provides an effectively unbiased reference against which the importance-weighting procedure can be validated.

To test for weighting biases, we modify the calibration pipeline by assigning every galaxy its COSMOS2020 redshift, $z_{\mathrm{C20}}$, regardless of whether a spectroscopic redshift is available. Setting $p_\mathrm{f, min} = 1$ corresponds to using the complete COSMOS2020 catalog with no importance weighting and provides a reference $n(z)$. Then we progressively decrease $p_\mathrm{f, min}$, applying the completeness weighting described in Section~\ref{sec:methods}, and measure whether the recovered $\bar{z}_{C20}$ remains consistent with the reference sample. The upper panels of Fig.~\ref{fig:d_z} show the shift in the weighted mean redshift, $\Delta\bar{z}_{\mathrm{C20}}$, relative to the unweighted case, while the lower panels provide an estimate of the statistical uncertainty on that difference.

For all surveys, the recovered mean redshift remains stable over a wide range of fiber fraction thresholds. For the fiducial choice of $p_\mathrm{f, min} = 0.1$, the shifts in $\bar{z}$ remain small, typically satisfying $\Delta\bar{z}_{\mathrm{C20}}<0.01$ for all tomographic bins. The lower panels exhibit a relatively flat and noisy behavior, with values spanning approximately $-3 \sigma \lesssim \Delta\bar{z}_{\mathrm{C20}} \lesssim 3 \sigma$. Statistical fluctuations are roughly on the scale of order unity and are expected. Overall, we find that the differences between the redshift distributions are mostly consistent with statistical fluctuations, and we see no evidence for systematic deviations exceeding $\sim 0.01$ in the mean redshift across the range of $p_{\mathrm{f}}$ considered.

Overall, these tests suggest that the importance-weighting approach provides an accurate reconstruction of redshift distributions despite significant spectroscopic incompleteness. Therefore we proceed with the fiducial choice $p_\mathrm{f, min} = 0.10$ in the remainder of this work as it provides a good compromise between accuracy, shot noise, and a high fraction of the calibration coming from spectroscopic redshifts. Current stage-III lensing surveys' mean redshifts are expected to be calibrated to within $\sigma_{\bar z} / (1 + \bar z) \sim 0.01$. The observed deviations stay well below that limit and advance toward the precision required for stage-IV surveys.

Finally, we note this validation tests the method's ability to recover intrinsic redshift distributions from biased samples. Additional systematic uncertainties, such as photometric calibration errors, are not explicitly addressed here.

\subsection{Redshift Sample}

\begin{figure*}
    \centering
    \includegraphics{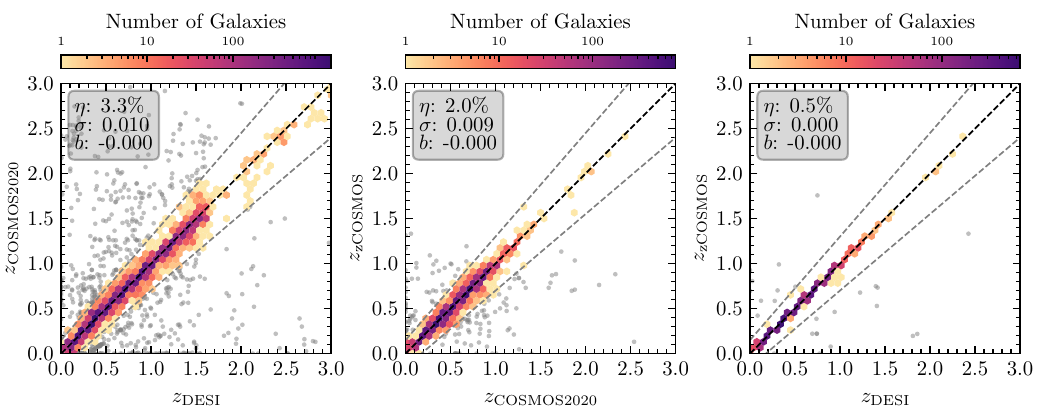}
    \caption{Comparison of redshifts computed by different surveys: spectroscopic surveys DESI and zCOSMOS, and photometric survey COSMOS2020. The black solid line corresponds to a one-to-one relation, and the dashed lines correspond to the redshift at $\pm0.15(1 + z)$. The fraction of sources outside the dashed lines (noted $\eta$), the precision measured with the normalized median absolute deviation (noted $\sigma$), and the overall bias (noted $b$) are indicated in the legend of each panel. Redshifts within the $\eta$ tolerance are binned on a $\mathrm{log}_{10}$ scale, and the outliers are indicated by single gray points.}
    \label{fig:redshift_quality}
\end{figure*}

We check the redshifts of each survey against one another, by plotting individual galaxy redshifts. In Fig.~\ref{fig:redshift_quality}, we see that the percentage of sources that lie outside the 15\% tolerance of a one-to-one relation is small, below 3.5\% for all comparison cases. This provides confidence that the three surveys recover the same or similar redshift per galaxy. We calculate the photo-z precision ($\sigma_{\mathrm{NMAD}}$) which stays below $0.01(1 + z)$. This formulation visually demonstrates the excellent agreement between catalogs.

\begin{table}
\centering
\caption{Photometric redshift calibration results comparison between the original pipeline and only using COSMOS2020 redshifts for the default parameter value, $p_\mathrm{f, min} = 0.10$.}
\begin{tabular}{ccc}
\hline
Sample & $\Delta z_\mathrm{C20 only}$ & $\sigma_{\Delta z}$ \\
\hline\hline
DECADE-1 & $0.004$ & $0.001$ \\
DECADE-2 & $0.007$ & $0.001$ \\
DECADE-3 & $0.001$ & $0.001$ \\
DECADE-4 & $0.000$ & $0.003$ \\
DES-1 & $0.003$ & $0.001$ \\
DES-2 & $0.007$ & $0.001$ \\
DES-3 & $0.001$ & $0.001$ \\
DES-4 & $0.003$ & $0.004$ \\
HSC-1 & $0.002$ & $0.001$ \\
HSC-2 & $-0.002$ & $0.001$ \\
HSC-3 & $-0.000$ & $0.001$ \\
HSC-4 & $0.001$ & $0.002$ \\
KiDS-1 & $0.001$ & $0.002$ \\
KiDS-2 & $-0.001$ & $0.001$ \\
KiDS-3 & $0.008$ & $0.003$ \\
KiDS-4 & $0.003$ & $0.003$ \\
KiDS-5 & $-0.000$ & $0.001$ \\
KiDS-6 & $0.006$ & $0.003$ \\
\hline
\end{tabular}
\tablecomments{$\Delta z_\mathrm{C20 only}$ is the difference mean redshift when only using COSMOS2020 redshifts. $\sigma_{\Delta z}$ is its associated uncertainty.}
\label{c2020_only}
\end{table}

As a verification of the validity of the COSMOS2020 photometric sample, we repeat the calibration using only the COSMOS2020 redshifts, excluding all spectroscopic information. This provides a test of whether the COSMOS2020 photometric sample alone produces redshift distributions consistent with those obtained from the full calibration sample. Table~\ref{c2020_only} shows the resulting redshift distribution means. We find the mean redshift differs between the COSMOS2020-only and unified calibrations by only a few thousandths, indicating the COSMOS2020 photometric redshifts provide a consistent representation of the $n(z)$ and supporting their use as the basis of the calibration pipeline.

\subsection{Neural Network Robustness Test with Gradient Boosting}

As a validation of the completeness correction, we test the robustness of the inferred fiber targeting probabilities by replacing the neural network model with an alternative machine learning approach based on gradient boosting.

We use the \texttt{HistGradientBoostingClassifier} implemented in \textsc{scikit-learn} to estimate the targeting probability $p_{\mathrm{f}}$. Gradient boosting is an ensemble method that constructs a sequence of decision trees, where each subsequent tree is trained to correct the residual errors of the previous ones. This approach provides a flexible model with different inductive biases, or the assumptions a model makes about how the inputs relate to outputs, compared to neural networks. We adopt the same input features and training labels as in the neural network case. Hyper-parameters are optimized through a grid search, varying the number of leaf nodes, learning rate, and feature subsampling fraction, while all other parameters are kept at their default values. The final model is trained using repeated $k$-fold cross-validation, following the same procedure as for the neural network, to obtain stable predictions of $p_{\mathrm{f}}$ and their uncertainties.

The performance of the gradient boosting model is comparable to that of the neural network. It achieves a similar $R^2$ score, with the neural network having only a marginal improvement. We then recompute the completeness weights and resulting redshift distributions using these alternative targeting probabilities. To quantify the impact on the calibration, we compare the mean redshift of each tomographic bin between the two methods. Uncertainties are estimated via bootstrap resampling. Across all surveys and bins, we find that shifts in the mean redshift are small, with $|\Delta \bar{z}| \lesssim 0.005$, and are typically significantly smaller than this threshold.

These results suggest that the inferred redshift distributions are robust to the choice of machine learning model used to estimate the targeting probabilities. The negligible differences show that systematic uncertainties associated with the completeness correction are minor compared to the statistical uncertainties of the calibration sample.

\section{Results}
\label{sec:results}

\begin{figure*}
    \centering
    \includegraphics{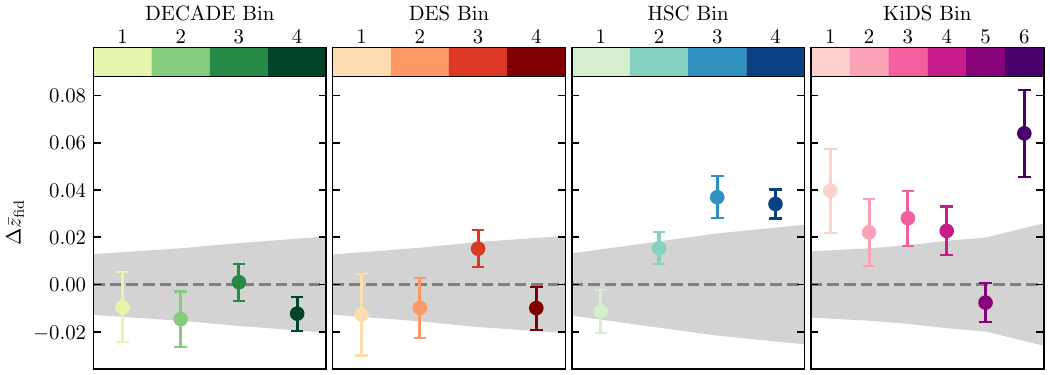}
    \caption{A one-dimensional plot of the results from Table~\ref{tab:n_z}. Fiducial value (zero) is shown in black, gray bars show the standard cosmological error value for photometric redshifts, colored points show the offset between this work and the fiducial, and error bars represent the shot noise, cosmic variance, where the bias is assumed to be 2, and redshift sample error summed in quadrature.}
    \label{fig:delta_z}
\end{figure*}

\begin{figure*}
    \centering
    \includegraphics{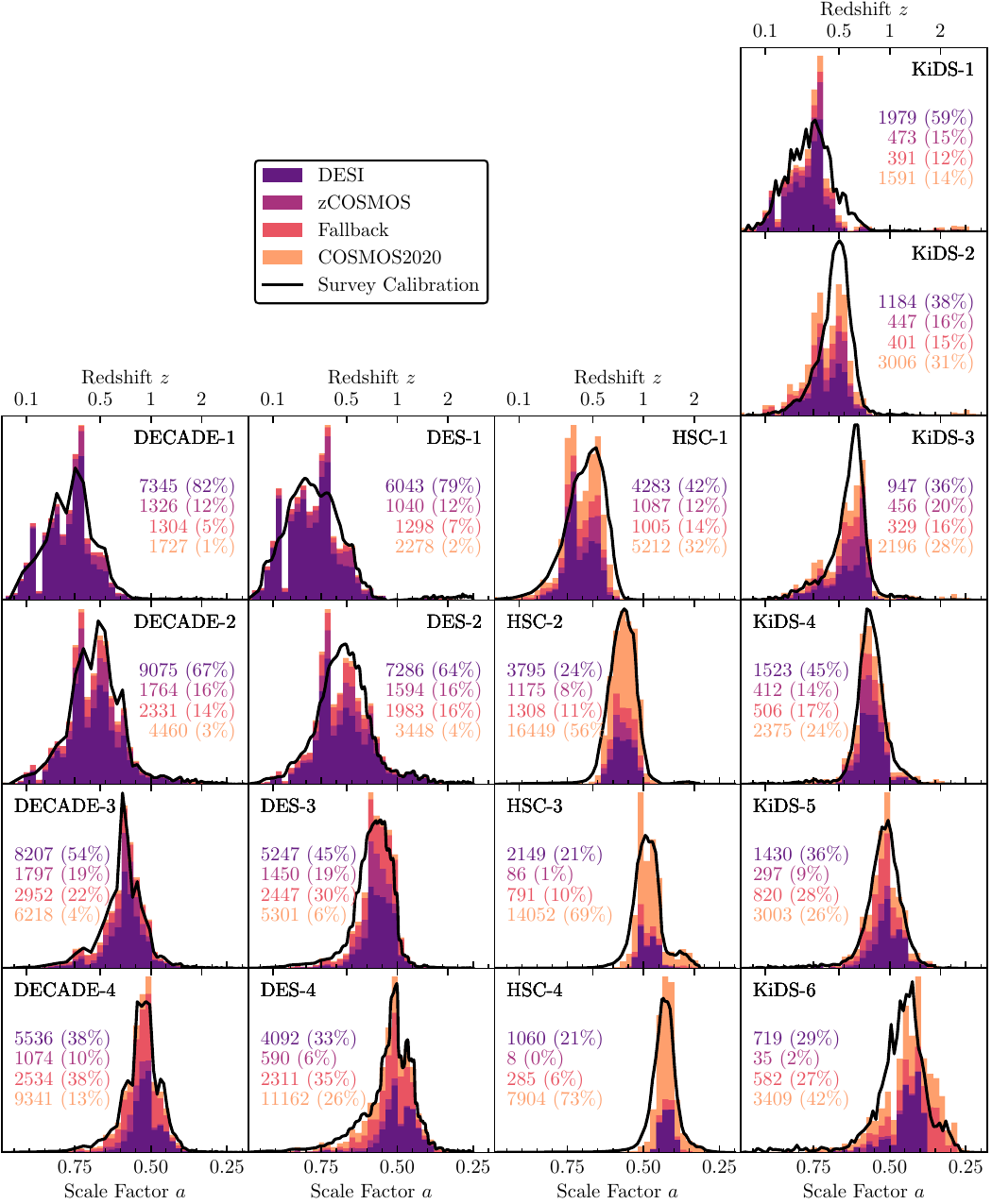}
    \caption{Effective redshift distributions, $n(z)$, estimated for the source galaxies in the tomographic bins of DECADE, DES Y3, HSC Y3, and KiDS-Legacy. The filled histograms present our calibrated distributions, with the galaxies separated and stacked according to redshift type. The legends give the number of galaxies in each category with its fractional contribution to the weighted $n(z)$. For reference, the fiducial distribution determined by each weak lensing collaboration is shown as a black line.}
    \label{fig:n_z}
\end{figure*}

\begin{table}
\centering
\caption{Photometric redshift calibration results for the default parameter value, $p_\mathrm{f, min} = 0.10$.}
\begin{tabular}{cccccc}
\hline
Sample & $\bar{z}$ & $\sigma_{\bar z, \mathrm{SN}}$ & $\sigma_{\bar z, \mathrm{CV}} / b$ & $\sigma_{\bar z, \mathrm{RS}}$ & $\Delta \bar{z}_\mathrm{fid}$ \\
\hline\hline
DECADE-1 & $0.324$ & $0.002$ & $0.013$ & $0.001$ & $-0.010$ \\
DECADE-2 & $0.488$ & $0.002$ & $0.011$ & $0.001$ & $-0.015$ \\
DECADE-3 & $0.708$ & $0.002$ & $0.007$ & $0.002$ & $+0.001$ \\
DECADE-4 & $0.896$ & $0.003$ & $0.005$ & $0.002$ & $-0.012$ \\
DES-1 & $0.324$ & $0.003$ & $0.016$ & $0.001$ & $-0.013$ \\
DES-2 & $0.512$ & $0.003$ & $0.011$ & $0.002$ & $-0.010$ \\
DES-3 & $0.756$ & $0.002$ & $0.006$ & $0.004$ & $+0.015$ \\
DES-4 & $0.926$ & $0.005$ & $0.007$ & $0.002$ & $-0.010$ \\
HSC-1 & $0.440$ & $0.002$ & $0.008$ & $0.002$ & $-0.011$ \\
HSC-2 & $0.782$ & $0.001$ & $0.006$ & $0.002$ & $+0.015$ \\
HSC-3 & $1.118$ & $0.002$ & $0.006$ & $0.005$ & $+0.037$ \\
HSC-4 & $1.364$ & $0.003$ & $0.004$ & $0.002$ & $+0.034$ \\
KiDS-1 & $0.400$ & $0.006$ & $0.015$ & $0.003$ & $+0.040$ \\
KiDS-2 & $0.486$ & $0.006$ & $0.011$ & $0.006$ & $+0.022$ \\
KiDS-3 & $0.617$ & $0.005$ & $0.009$ & $0.002$ & $+0.028$ \\
KiDS-4 & $0.804$ & $0.004$ & $0.006$ & $0.006$ & $+0.023$ \\
KiDS-5 & $0.943$ & $0.004$ & $0.006$ & $0.004$ & $-0.008$ \\
KiDS-6 & $1.343$ & $0.009$ & $0.008$ & $0.013$ & $+0.064$ \\
\hline
\end{tabular}
\tablecomments{The rightmost column shows the difference with respect to the fiducial results published by the weak lensing collaborations.}
\label{tab:n_z}
\end{table}

Fig.~\ref{fig:delta_z} presents a one-dimensional visualization of our $\Delta \bar z_{\mathrm{fid}}$ shifts for DECADE, DES Y3, HSC Y3, and KiDS-Legacy, and Fig.~\ref{fig:n_z} presents our estimates of the effective redshift distributions. We compare these results to the fiducial $n(z)$ published by the respective survey collaborations \citep{Myles2021_MNRAS_505_4249, Rau2023_MNRAS_524_5109, Wright2025_AA_703_144, Anbajagane2025_OJAp_8_6159}. For visualization, we show the distributions as a function of scale factor $a = (1+z)^{-1}$, which compresses the high-redshift tail and highlights the differences between methods.  Table~\ref{tab:n_z} summarizes the mean redshifts $\bar z$ for each tomographic bin, including estimates of shot noise, cosmic variance, and redshift sample uncertainty, as well as the differences relative to the fiducial results published by the weak lensing collaborations.

\subsection{Dark Energy Camera All Data Everywhere}

We find broad agreement with the DECADE fiducial analysis \citep{Anbajagane2025_OJAp_8_6159}. The mean redshifts quoted here are based on our direct calibration and do not include correction for blending effects, which are also unaccounted for in the fiducial analysis. The overall agreement is similar to the agreement found for DES Y3, likely owing to similarities in the calibration methodologies of the two surveys. We find a mean absolute offset of $\sim 0.01$ in the mean redshifts relative to the fiducial analysis.

\subsection{Dark Energy Survey}

After accounting for blending, we find broad agreement with the \citet{Myles2021_MNRAS_505_4249} DES Y3 calibration, both in terms of the mean and distributions of redshifts. The fiducial DES Y3 analysis accounts for blending effects using image simulations, which modify the $n(z)$ relevant for the lensing measurement \citep{MacCrann2022_MNRAS_509_3371}. Therefore, we incorporate the corresponding blending corrections into our $\bar z$ estimates when comparing with the fiducial results. The largest shift is observed in the third tomographic bin (DES-3), with $\Delta \bar z_\mathrm{fid} = +0.015$, while the remaining bins show smaller deviations consistent with zero.

\subsection{Hyper Suprime-Cam}

For HSC Y3, we observe good agreement in the lower-redshift tomographic bins, while the higher-redshift bins exhibit more noticeable differences with respect to the fiducial results \citep{Rau2023_MNRAS_524_5109}. We see a higher mean redshift for the two highest bins, with offsets of $+0.037 \pm 0.013$ and $+0.034 \pm 0.009$, respectively, where the errors are shot noise, covariance, and redshift sample error summed in quadrature and the galaxy bias $b_\mathrm{gal}$ is conservatively assumed to be 2. In these two bins, we observe that the bulk of the redshift distribution is similar to the fiducial analysis, however the black comparison distribution line suggests the presence of a lower redshift tail that is not present in our analysis. 

However, our results agree well with previously published independent results that calibrate via clustering methods using DESI DR1 and DR2 \citep{ChoppindeJanvry2026_JCAP_05_073}. This result finds offsets of $+0.043$ and $+0.050$ in the two highest respective bins, an absolute difference of $0.006$ and $0.016$ with respect to our direct calibration results.

\subsection{Kilo-Degree Survey}

For KiDS-Legacy, we find more pronounced differences relative to the published calibration \citep{Wright2025_AA_703_144}, similar to the re-analysis of KiDS-1000 \citep{Hildebrandt2021_AA_647_124} by \cite{Lange2026_ApJ_998_263}. In all tomographic bins excluding KiDS-5, our results indicate an increased contribution from galaxies at higher redshift compared to the fiducial $n(z)$. At the same time, the bulk of the distributions in KiDS-1 through 4 is shifted toward slightly lower redshifts, with the mean remaining relatively stable due to extended high-redshift tails.

\section{Discussion}
\label{sec:discussion}

\subsection{Inclusion of zCOSMOS-bright}

The motivation behind the addition of zCOSMOS-bright was to improve coverage of the galaxy population in regions of color-space that are sparsely sampled by DESI alone, and to cross-check potential systematic biases associated with the usage of only one spectroscopic training sample. We find that the inclusion of zCOSMOS improves the number of galaxies targeted by spectroscopy, especially in color spaces that are underrepresented by DESI alone. However, due to the much lower success rate of zCOSMOS, the overall weighted fraction of the weak lensing samples calibrated with spectroscopic redshifts remains essentially unchanged. For example, for DES Y3, the overall fraction of the $n(z)$ calibrated with spectroscopic redshifts changes by $-2\%$, $+2\%$, $+2\%$, and $-3\%$ for the four tomographic bins compared to \citet{Lange2026_ApJ_998_263}. That is because the fraction of galaxies that were spectroscopically targeted but did not end up getting a secure redshift increased from $3\%$, $4\%$, $3\%$, and $5\%$ to $7\%$, $16\%$, $30\%$ and $35\%$. This highlights the need for spectroscopic reference samples with high redshift success. We remind the reader that only selecting galaxies with secure redshifts could easily bias the redshift distributions if the redshift success rate depends directly or indirectly on the redshift itself \citep{Hartley2020_MNRAS_496_4769}.

\subsection{Cosmology Implications for HSC Y3}

\begin{figure}
    \centering
    \includegraphics{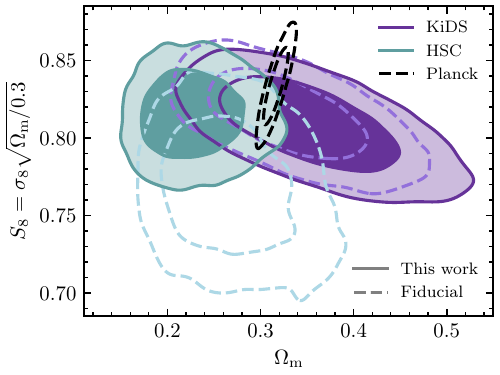}
    \caption{Posterior constraints on $S_8$ and $\Omega_m$ from the KiDS-Legacy and HSC Y3 cosmic shear analyses. In dashed lines, we show the fiducial results from \cite{Wright2025_AA_703_158} and \cite{Dalal2023_PhRvD_108_3519}, whereas the updated results using the redshift distribution derived in this work are shown in solid. The innermost contour represents the $1\sigma$ confidence interval, and the outermost, $2\sigma$. The \cite{PlanckCollaboration2020_AA_641_6} TT,TE,EE+lowE results in black.}
    \label{fig:s8}
\end{figure}

For HSC Y3, our direct unified calibration provides an independent assessment of the tomographic redshift distributions used in the HSC Y3 cosmic shear analysis \citep{Dalal2023_PhRvD_108_3519, Li2022_PASJ_74_421}. The comparison is especially relevant for the two highest redshift bins, where the fiducial HSC calibration is less directly constrained.

In the HSC Y3 fiducial analysis, photometric redshift information is combined with spatial cross-correlations between the HSC source sample and the CAMIRA Luminous Red Galaxy (LRG) sample \citep{Rau2023_MNRAS_524_5109}. The CAMIRA LRG sample extends to $z<1.2$, providing clustering-redshift information for bins 1 and 2 and partial coverage of bin 3 ($0.9<z\le1.2$). The fourth tomographic bin ($1.2<z\leq1.5$) has no corresponding cross-correlation data vector. In addition to the limited range, the LRG sample redshifts are photometric and therefore subject to redshift uncertainty \citep{Rau2023_MNRAS_524_5109}.

In the subsequent HSC Y3 cosmic shear analysis, uncertainty in the source redshift distributions is parameterized using the nuisance parameter, $\Delta z_i$, which represents an uncertainty in $n_i(z)$ as a free shift in $\bar z_i$ during cosmological inference. For the third and fourth bins, \citet{Dalal2023_PhRvD_108_3519} adopted a wide, flat prior given that these bins are only partially or not at all constrained by the CAMIRA LRG clustering calibration. This choice is also informed by $S_8$ being shown to shift to significantly higher values when both Gaussian priors are assumed on all bins and the fourth tomographic bin is excluded, which is suspected to be due to bins 3 and 4 being systematically biased by the three neural network-based conditional density estimation codes \citep[Section V E of][]{Dalal2023_PhRvD_108_3519}. The conservative posterior constraints from the cosmic shear likelihood correspond to $\Delta z_3=+0.076^{+0.059}_{-0.056}$ and $\Delta z_4=+0.157^{+0.111}_{-0.094}$, where positive $\Delta z_i$ denotes a shift toward higher redshifts. 

Our direct calibration method finds $\bar z$ shifts of $-0.011$ and $+0.015$ for bins 1 and 2, indicating a broad agreement with the HSC fiducial analysis including taking into account uncertainties. Summing shot noise, covariance, and redshift sample error in quadrature, our total uncertainty is $\pm 0.016$ in bin 1 and $\pm 0.012$ in bin 2. HSC reports their fiducial uncertainties for bins 1 and 2 as $0.024$ and $0.022$, respectively, indicating that our inferred offsets for these bins are consistent with their results within errors. In contrast, we find disagreement in the third and fourth bins, where fiducial LRG coverage is the sparsest with $\Delta \bar z_3=+0.037$ and $\Delta \bar z_4=+0.034$, respectively. These shifts are substantially smaller than those preferred by the HSC cosmic shear likelihood, but closely agree with independent DESI clustering calibration from \citet{ChoppindeJanvry2026_JCAP_05_073}. Specifically, this analysis finds shifts of approximately $+0.043$ and $+0.050$ in bins 3 and 4, respectively. This similarity is notable due to the different methods used to constrain the source redshift distributions.

The difference between these externally calibrated shifts and those preferred by the HSC cosmic shear likelihood has direct implications for the interpretation of HSC Y3 cosmological constraints. The HSC Y3 cosmic shear analysis finds a lower value of $S_8$ when wide, flat priors are used with $S_8=0.776^{+0.033}_{-0.034}$ \citep{Dalal2023_PhRvD_108_3519, Rana2026_PhRvD_113_3534}. When more informative Gaussian priors recommended by \citet{Rau2023_MNRAS_524_5109} are used, the inferred value increases to $S_8=0.813^{+0.024}_{-0.025}$, while the error decreases substantially. This broadly agrees with our inferred $S_8$ value of $0.815^{+0.018}_{-0.018}$ and tighter constraints shown in Fig.~\ref{fig:s8} where we assume priors on the mean redshift biases by summing the shot noise, cosmic variance, and redshift sample uncertainties in quadrature, assuming a galaxy bias of $2$. Our results also agree with independent cosmic shear results \citep{ChoppindeJanvry2025_arXiv_2511_8134}.  \footnote{The HSC fiducial posterior is obtained via the \textsc{PolyChord} nested sampler \citep{Handley2015_MNRAS_453_4384} as described in \citet{Li2023_PhRvD_108_3518}. Our posterior is produced using methods described in \citet{ChoppindeJanvry2025_arXiv_2511_8134}, which uses the \textsc{cosmoSIS} pipeline \citep{Zuntz2015_AC_12_45} and the \textsc{pocoMC} Preconditioned Monte Carlo sampler \citep{Karamanis2022_MNRAS_516_1644, Karamanis2022_JOSS_7_4634}.}

\subsection{Cosmology Implications for KiDS-Legacy}

In contrast to DES and DECADE, we find more pronounced differences with respect to the KiDS-Legacy redshift distributions. Our calibration yields high-redshift tails in several tomographic bins, while maintaining similar mean redshifts in some cases due to compensating shifts in the bulk of the distribution. These differences could translate into shifts in the inferred cosmological parameters, such as through changes in the lensing amplitude. For example, an increase in the mean redshift of the source distribution generally leads to a lower inferred value of $S_8$, potentially increasing tension with the Planck CMB results \citep{PlanckCollaboration2020_AA_641_6}. To test this, we repeat the cosmic shear analysis carried out in \citet{Wright2025_AA_703_158} using our updated $n(z)$ estimates. Ultimately, when comparing the inferred $S_8$--$\Omega_m$ constraints in Fig.~\ref{fig:s8}, we do not observe a significant shift relative to the KiDS-Legacy fiducial calibration, finding $S_8 = 0.808_{-0.020}^{+0.019}$ compared to the fiducial result of $S_8 = 0.815_{-0.019}^{+0.017}$. Moreover, the shift is primarily due to a shift to higher $\Omega_\mathrm{m}$ while the preferred $S_8$ at the Planck-preferred value of $\Omega_\mathrm{m}$ is essentially unchanged. \footnote{The KiDS-Legacy posteriors in Fig.~\ref{fig:s8} are obtained using \textsc{nautilus}, a Bayesian posterior and evidence estimation code utilizing importance sampling and neural networks \citep{Lange2023_MNRAS_525_3181}.}

\subsection{DES Y6 Redshift Sample Error}

The redshift sample error is a useful point of comparison with the recent DES Y6 analysis \citep{Yin2025_arXiv_2510_3566}. For the DES Y3 source sample, we find $\sigma_{\bar z, \mathrm{RS}}=0.001, 0.002, 0.004, 0.002$ in tomographic bins 1--4. \citet{Yin2025_arXiv_2510_3566} report a substantially larger contribution from the redshift calibration sample for Y6, with total mean-redshift uncertainties of $0.012, 0.008, 0.009, 0.024$, of which the redshift sample term is the dominant component. Part of this difference is expected. The Y6 source sample extends roughly one magnitude deeper ($i < 24.7$ versus $i<23.5$), and as a result, it relies far more heavily on photometric redshifts. The spectroscopic fraction of the calibration sample in Y6 is $48\%, 41\%, 40\%,$ and $19\%$ in bins 1--4 \citep[see Fig. 3 of][]{Yin2025_arXiv_2510_3566}, whereas in our calibration of Y3, it is $90\%, 80\%, 64\%,$ and $39\%$ (of which $79\%, 64\%, 45\%,$ and $33\%$ come from DESI). A larger photometric fraction naturally increases the sensitivity to COSMOS2020 redshift biases. The two analyses also differ in the modeling of that bias. Y6 uses spectroscopic comparisons binned in magnitude and redshift to perturb the COSMOS2020 and PAU redshifts, whereas we use the offset between the \textsc{EAZY} and \textsc{LePhare} redshifts. Therefore, we do not expect the two error budgets to agree exactly, and we do not attempt to attribute the difference to any one choice. A more detailed comparison is left to future work.

\section{Conclusion}
\label{sec:conclusion}

In this work, we have extended the first unified photometric redshift calibration framework introduced in \citet{Lange2026_ApJ_998_263} to incorporate updated photometric data sets and an expanded spectroscopic calibration sample. We replace earlier survey releases with HSC Y3 and KiDS-Legacy, add DECADE, and add the zCOSMOS spectroscopic sample as an additional tier in the calibration pipeline.

Applying this framework to the four surveys we find:

\begin{itemize}
    \item The addition of zCOSMOS, while increasing the number of spectroscopic redshifts used, does not improve the weighted fraction of galaxies calibrated with spectroscopy, owing to the increased redshift failure rate of zCOSMOS-bright compared to DESI.
    \item We find broad agreement with the DECADE and DES Y3 fiducial results.
    \item We do not find evidence for the strong shifts proposed in the \citet{Li2023_PhRvD_108_3518, Dalal2023_PhRvD_108_3519} analyses for HSC Y3, leading to a higher inferred value of $S_8$ and a significant reduction in the $S_8$-tension.
    \item While we do find significant differences in the redshift distribution for KiDS-Legacy, our final cosmological constraints appear to have minimal impact on the (absence of an) $S_8$-tension in this data set.
    \item We find that redshift sampling uncertainty does not significantly affect our analysis, however we leave a deeper investigation of its impact to future work.
\end{itemize}

These results demonstrate the unified calibration framework provides a robust and largely independent cross-check of current photometric redshift calibration methods. This method complements existing SOM- and clustering-based approaches by relying on a reconstruction of $n(z)$ using diverse spectroscopic samples and an importance weighting framework. Looking ahead, improving photometric redshift calibration will require spectroscopic samples with high completeness and success rates across the color--magnitude space of weak lensing samples. Deeper spectroscopy in fields such as COSMOS and XMM and high-precision photometry from surveys such as LSST and Roman will provide better coverage of the faint galaxy populations that currently limit calibration.

Our direct calibration method provides an independent approach to assessing the redshift distributions used in weak lensing analyses. In particular, we will apply this framework to the DESI DR2 3$\times$2pt analysis where accurate photometric redshift distributions are needed to jointly constrain cosmology using galaxy clustering, galaxy--galaxy lensing, and cosmic shear \citep{Dey2026_AJ_172_65, DeRose2026_arXiv_2603_0113}.

\section*{Data Availability}

All data points shown in the published graphs will be available at https://zenodo.org/records/24086052 upon publication.

\section*{Acknowledgments}

We thank Angus Wright for the production of the KiDS-Legacy cosmological posterior shown in Fig.~\ref{fig:s8}, and his and Justin Myles' role in the discussions surrounding this analysis. \textsc{Claude Sonnet 5.5} was used to check parts of the code used in this analysis for inaccuracies and inconsistencies with the text.

This work made use of the following software: \textsc{Astropy} \citep{AstropyCollaboration2013_AA_558_33, AstropyCollaboration2018_AJ_156_123, AstropyCollaboration2022_ApJ_935_167}, \textsc{matplotlib} \citep{Hunter2007_CSE_9_90}, \textsc{NumPy} \citep{VanDerWalt2011_CSE_13_22}, \textsc{SciPy} \citep{Virtanen2020_NaMet_17_261}, \textsc{scikit-learn} \citep{Pedregosa2011_JMLR_12_2825}, \textsc{cosmoSIS} \citep[][https://github.com/cosmosis-developers/cosmosis/]{Zuntz2015_AC_12_45}, \textsc{pocoMC} \citep{Karamanis2022_MNRAS_516_1644, Karamanis2022_JOSS_7_4634}, and \textsc{Spyder} (https://github.com/kwaldenphd/python-spyder/).

\textbf{Dark Energy Spectroscopic Instrument:} This material is based upon work supported by the U.S. Department of Energy (DOE), Office of Science, Office of High-Energy Physics, under Contract No. DE–AC02–05CH11231, and by the National Energy Research Scientific Computing Center, a DOE Office of Science User Facility under the same contract. Additional support for DESI was provided by the U.S. National Science Foundation (NSF), Division of Astronomical Sciences under Contract No. AST-0950945 to the NSF’s National Optical-Infrared Astronomy Research Laboratory; the Science and Technology Facilities Council of the United Kingdom; the Gordon and Betty Moore Foundation; the Heising-Simons Foundation; the French Alternative Energies and Atomic Energy Commission (CEA); the Secretariat of Science, Humanities, Technology and Innovation (SECIHTI) of Mexico; the Ministry of Science, Innovation and Universities of Spain (MICIU/AEI/10.13039/501100011033), and by the DESI Member Institutions: \url{https://www.desi.lbl.gov/collaborating-institutions}.

The DESI Legacy Imaging Surveys consist of three individual and complementary projects: the Dark Energy Camera Legacy Survey (DECaLS), the Beijing-Arizona Sky Survey (BASS), and the Mayall z-band Legacy Survey (MzLS). DECaLS, BASS and MzLS together include data obtained, respectively, at the Blanco telescope, Cerro Tololo Inter-American Observatory, NSF’s NOIRLab; the Bok telescope, Steward Observatory, University of Arizona; and the Mayall telescope, Kitt Peak National Observatory, NOIRLab. NOIRLab is operated by the Association of Universities for Research in Astronomy (AURA) under a cooperative agreement with the National Science Foundation. Pipeline processing and analyses of the data were supported by NOIRLab and the Lawrence Berkeley National Laboratory. Legacy Surveys also uses data products from the Near-Earth Object Wide-field Infrared Survey Explorer (NEOWISE), a project of the Jet Propulsion Laboratory/California Institute of Technology, funded by the National Aeronautics and Space Administration. Legacy Surveys was supported by: the Director, Office of Science, Office of High Energy Physics of the U.S. Department of Energy; the National Energy Research Scientific Computing Center, a DOE Office of Science User Facility; the U.S. National Science Foundation, Division of Astronomical Sciences; the National Astronomical Observatories of China, the Chinese Academy of Sciences and the Chinese National Natural Science Foundation. LBNL is managed by the Regents of the University of California under contract to the U.S. Department of Energy. The complete acknowledgments can be found at \url{https://www.legacysurvey.org/}.

Any opinions, findings, and conclusions or recommendations expressed in this material are those of the author(s) and do not necessarily reflect the views of the U. S. National Science Foundation, the U. S. Department of Energy, or any of the listed funding agencies.

The authors are honored to be permitted to conduct scientific research on I'oligam Du'ag (Kitt Peak), a mountain with particular significance to the Tohono O’odham Nation.

\textbf{Dark Energy Camera All Data Everywhere:} The DELVE Collaboration, which processed the DECADE dataset used in this work, gratefully acknowledges support from Fermilab LDRD L2019-011, the NASA Fermi Guest Investigator Program Cycle 9 No. 91201, and the National Science Foundation under Grant No. AST-2108168, AST-2108169, AST-2307126, and AST-2407526. The DELVE research is partially funded by a generous gift from Charles Simonyi to the NSF Division of Astronomical Sciences. The award is made in recognition of significant contributions to Rubin Observatory's Legacy Survey of Space and Time.

\textbf{Dark Energy Survey:} This project used public archival data from the Dark Energy Survey (DES). Funding for the DES Projects has been provided by the U.S. Department of Energy, the U.S. National Science Foundation, the Ministry of Science and Education of Spain, the Science and Technology Facilities Council of the United Kingdom, the Higher Education Funding Council for England, the National Center for Supercomputing Applications at the University of Illinois at Urbana-Champaign, the Kavli Institute of Cosmological Physics at the University of Chicago, the Center for Cosmology and Astro-Particle Physics at the Ohio State University, the Mitchell Institute for Fundamental Physics and Astronomy at Texas A\&M University, Financiadora de Estudos e Projetos, Funda{\c c}{\~a}o Carlos Chagas Filho de Amparo {\`a} Pesquisa do Estado do Rio de Janeiro, Conselho Nacional de Desenvolvimento Cient{\'i}fico e Tecnol{\'o}gico and the Minist{\'e}rio da Ci{\^e}ncia, Tecnologia e Inova{\c c}{\~a}o, the Deutsche Forschungsgemeinschaft, and the Collaborating Institutions in the Dark Energy Survey.

The Collaborating Institutions are Argonne National Laboratory, the University of California at Santa Cruz, the University of Cambridge, Centro de Investigaciones Energ{\'e}ticas, Medioambientales y Tecnol{\'o}gicas-Madrid, the University of Chicago, University College London, the DES-Brazil Consortium, the University of Edinburgh, the Eidgen{\"o}ssische Technische Hochschule (ETH) Z{\"u}rich,  Fermi National Accelerator Laboratory, the University of Illinois at Urbana-Champaign, the Institut de Ci{\`e}ncies de l'Espai (IEEC/CSIC), the Institut de F{\'i}sica d'Altes Energies, Lawrence Berkeley National Laboratory, the Ludwig-Maximilians Universit{\"a}t M{\"u}nchen and the associated Excellence Cluster Universe, the University of Michigan, the National Optical Astronomy Observatory, the University of Nottingham, The Ohio State University, the OzDES Membership Consortium, the University of Pennsylvania, the University of Portsmouth, SLAC National Accelerator Laboratory, Stanford University, the University of Sussex, and Texas A\&M University.

Based in part on observations at Cerro Tololo Inter-American Observatory, National Optical Astronomy Observatory, which is operated by the Association of Universities for Research in Astronomy (AURA) under a cooperative agreement with the National Science Foundation.

\textbf{Hyper Suprime Cam:} The Hyper Suprime-Cam (HSC) collaboration includes the astronomical communities of Japan and Taiwan, and Princeton University. The HSC instrumentation and software were developed by the National Astronomical Observatory of Japan (NAOJ), the Kavli Institute for the Physics and Mathematics of the Universe (Kavli IPMU), the University of Tokyo, the High Energy Accelerator Research Organization (KEK), the Academia Sinica Institute for Astronomy and Astrophysics in Taiwan (ASIAA), and Princeton University. Funding was contributed by the FIRST program from the Japanese Cabinet Office, the Ministry of Education, Culture, Sports, Science and Technology (MEXT), the Japan Society for the Promotion of Science (JSPS), Japan Science and Technology Agency (JST), the Toray Science Foundation, NAOJ, Kavli IPMU, KEK, ASIAA, and Princeton University. 

This paper makes use of software developed for Vera C. Rubin Observatory. We thank the Rubin Observatory for making their code available as free software at \url{http://pipelines.lsst.io/}.

This paper is based on data collected at the Subaru Telescope and retrieved from the HSC data archive system, which is operated by the Subaru Telescope and Astronomy Data Center (ADC) at NAOJ. Data analysis was in part carried out with the cooperation of Center for Computational Astrophysics (CfCA), NAOJ. We are honored and grateful for the opportunity of observing the Universe from Maunakea, which has the cultural, historical and natural significance in Hawaii. 

The Pan-STARRS1 Surveys (PS1) and the PS1 public science archive have been made possible through contributions by the Institute for Astronomy, the University of Hawaii, the Pan-STARRS Project Office, the Max Planck Society and its participating institutes, the Max Planck Institute for Astronomy, Heidelberg, and the Max Planck Institute for Extraterrestrial Physics, Garching, The Johns Hopkins University, Durham University, the University of Edinburgh, the Queen’s University Belfast, the Harvard-Smithsonian Center for Astrophysics, the Las Cumbres Observatory Global Telescope Network Incorporated, the National Central University of Taiwan, the Space Telescope Science Institute, the National Aeronautics and Space Administration under grant No. NNX08AR22G issued through the Planetary Science Division of the NASA Science Mission Directorate, the National Science Foundation grant No. AST-1238877, the University of Maryland, Eotvos Lorand University (ELTE), the Los Alamos National Laboratory, and the Gordon and Betty Moore Foundation.

\textbf{Kilo-Degree Survey:} Based on data obtained from the ESO Science Archive Facility with DOI: \url{https://doi.org/10.18727/archive/37}, and \url{https://doi.eso.org/10.18727/archive/59} and on data products produced by the KiDS consortium. The KiDS production team acknowledges support from: Deutsche Forschungsgemeinschaft, ERC, NOVA and NWO-M grants; Target; the University of Padova, and the University Federico II (Naples).

\bibliographystyle{mnras}
\bibliography{bibliography}

\section*{Affiliations}
\noindent
{\footnotesize $^{1}$ Department of Physics, American University, 4400 Massachusetts Avenue NW, Washington, DC 20016, USA\\
$^{2}$ Centre for Astrophysics \& Supercomputing, Swinburne University of Technology, P.O. Box 218, Hawthorn, VIC 3122, Australia\\
$^{3}$ Center for Cosmology and AstroParticle Physics, The Ohio State University, 191 West Woodruff Avenue, Columbus, OH 43210, USA\\
$^{4}$ Department of Astronomy, The Ohio State University, 4055 McPherson Laboratory, 140 W 18th Avenue, Columbus, OH 43210, USA\\
$^{5}$ Department of Physics, The Ohio State University, 191 West Woodruff Avenue, Columbus, OH 43210, USA\\
$^{6}$ The Ohio State University, Columbus, 43210 OH, USA\\
$^{7}$ Department of Astronomy, Cornell University, Ithaca, NY 14853, USA\\
$^{8}$ Department of Physics, University of California, Berkeley, CA 94720, USA\\
$^{9}$ Lawrence Berkeley National Laboratory, 1 Cyclotron Road, Berkeley, CA 94720, USA\\
$^{10}$ Berkeley Center for Cosmological Physics, Department of Physics, University of California, Berkeley, CA 94720, USA\\
$^{11}$ Department of Astronomy and Astrophysics, UCO/Lick Observatory, University of California, 1156 High Street, Santa Cruz, CA 95064, USA\\
$^{12}$ Department of Physics and Astronomy, University of Waterloo, 200 University Ave W, Waterloo, ON N2L 3G1, Canada\\
$^{13}$ Perimeter Institute for Theoretical Physics, 31 Caroline St. North, Waterloo, ON N2L 2Y5, Canada\\
$^{14}$ Waterloo Centre for Astrophysics, University of Waterloo, 200 University Ave W, Waterloo, ON N2L 3G1, Canada\\
$^{15}$ CIEMAT, Avenida Complutense 40, E-28040 Madrid, Spain\\
$^{16}$ Institute for Astronomy, University of Edinburgh, Royal Observatory, Blackford Hill, Edinburgh EH9 3HJ, UK\\
$^{17}$ Ruhr University Bochum, Faculty of Physics and Astronomy, Astronomical Institute (AIRUB), German Centre for Cosmological Lensing, 44780 Bochum, Germany\\
$^{18}$ Queensland University of Technology,  School of Chemistry \& Physics, George St, Brisbane 4001, Australia\\
$^{19}$ Department of Astronomy, School of Physics and Astronomy, Shanghai Jiao Tong University, Shanghai 200240, China\\
$^{20}$ University Observatory, Faculty of Physics, Ludwig-Maximilians-Universit\"{a}t, Scheinerstr. 1, 81679 M\"{u}nchen, Germany\\
$^{21}$ Department of Astronomy, Tsinghua University, 30 Shuangqing Road, Haidian District, Beijing, China, 100190\\
$^{22}$ Department of Physics, Boston University, 590 Commonwealth Avenue, Boston, MA 02215 USA\\
$^{23}$ Instituto Avanzado de Cosmolog\'{\i}a A.~C., San Marcos 11 - Atenas 202. Magdalena Contreras. Ciudad de M\'{e}xico C.~P.~10720, M\'{e}xico\\
$^{24}$ Instituto de Ciencias F\'{\i}sicas, Universidad Nacional Aut\'onoma de M\'exico, Av. Universidad s/n, Cuernavaca, Morelos, C.~P.~62210, M\'exico\\
$^{25}$ Dipartimento di Fisica ``Aldo Pontremoli'', Universit\`a degli Studi di Milano, Via Celoria 16, I-20133 Milano, Italy\\
$^{26}$ INAF-Osservatorio Astronomico di Brera, Via Brera 28, 20122 Milano, Italy\\
$^{27}$ Department of Physics \& Astronomy, University College London, Gower Street, London, WC1E 6BT, UK\\
$^{28}$ Departamento de Astrof\'{\i}sica, Universidad de La Laguna (ULL), E-38206, La Laguna, Tenerife, Spain\\
$^{29}$ Instituto de Astrof\'{\i}sica de Canarias, C/ V\'{\i}a L\'{a}ctea, s/n, E-38205 La Laguna, Tenerife, Spain\\
$^{30}$ Institut d'Estudis Espacials de Catalunya (IEEC), c/ Esteve Terradas 1, Edifici RDIT, Campus PMT-UPC, 08860 Castelldefels, Spain\\
$^{31}$ Institute of Space Sciences, ICE-CSIC, Campus UAB, Carrer de Can Magrans s/n, 08913 Bellaterra, Barcelona, Spain\\
$^{32}$ Instituto de F\'{\i}sica, Universidad Nacional Aut\'{o}noma de M\'{e}xico,  Circuito de la Investigaci\'{o}n Cient\'{\i}fica, Ciudad Universitaria, Cd. de M\'{e}xico  C.~P.~04510,  M\'{e}xico\\
$^{33}$ Department of Astronomy \& Astrophysics, University of Toronto, Toronto, ON M5S 3H4, Canada\\
$^{34}$ Department of Physics \& Astronomy and Pittsburgh Particle Physics, Astrophysics, and Cosmology Center (PITT PACC), University of Pittsburgh, 3941 O'Hara Street, Pittsburgh, PA 15260, USA\\
$^{35}$ Instituci\'{o} Catalana de Recerca i Estudis Avan\c{c}ats, Passeig de Llu\'{\i}s Companys, 23, 08010 Barcelona, Spain\\
$^{36}$ Institut de F\'{i}sica d’Altes Energies (IFAE), The Barcelona Institute of Science and Technology, Edifici Cn, Campus UAB, 08193 Bellaterra (Barcelona), Spain\\
$^{37}$ Departamento de F\'isica, Universidad de los Andes, Cra. 1 No. 18A-10, Edificio Ip, CP 111711, Bogot\'a, Colombia\\
$^{38}$ Observatorio Astron\'omico, Universidad de los Andes, Cra. 1 No. 18A-10, Edificio H, CP 111711 Bogot\'a, Colombia\\
$^{39}$ Institute of Cosmology and Gravitation, University of Portsmouth, Dennis Sciama Building, Portsmouth, PO1 3FX, UK\\
$^{40}$ University of Virginia, Department of Astronomy, Charlottesville, VA 22904, USA\\
$^{41}$ Fermi National Accelerator Laboratory, PO Box 500, Batavia, IL 60510, USA\\
$^{42}$ Institute of Astronomy, University of Cambridge, Madingley Road, Cambridge CB3 0HA, UK\\
$^{43}$ University of California, Berkeley, 110 Sproul Hall \#5800 Berkeley, CA 94720, USA\\
$^{44}$ Department of Astronomy, University of Texas at Austin, 2515 Speedway, TX 78712, USA\\
$^{45}$ Institut d'Astrophysique de Paris. 98 bis boulevard Arago. 75014 Paris, France\\
$^{46}$ IRFU, CEA, Universit\'{e} Paris-Saclay, F-91191 Gif-sur-Yvette, France\\
$^{47}$ NSF NOIRLab, 950 N. Cherry Ave., Tucson, AZ 85719, USA\\
$^{48}$ Department of Physics and Astronomy, University of California, Irvine, 92697, USA\\
$^{49}$ Sorbonne Universit\'{e}, CNRS/IN2P3, Laboratoire de Physique Nucl\'{e}aire et de Hautes Energies (LPNHE), FR-75005 Paris, France\\
$^{50}$ Department of Astronomy and Astrophysics, University of California, Santa Cruz, 1156 High Street, Santa Cruz, CA 95065, USA\\
$^{51}$ Departament de F\'{i}sica, Serra H\'{u}nter, Universitat Aut\`{o}noma de Barcelona, 08193 Bellaterra (Barcelona), Spain\\
$^{52}$ Department of Physics and Astronomy, Siena University, 515 Loudon Road, Loudonville, NY 12211, USA\\
$^{53}$ Department of Physics \& Astronomy, University  of Wyoming, 1000 E. University, Dept.~3905, Laramie, WY 82071, USA\\
$^{54}$ Instituto de Estudios Astrof\'isicos, Facultad de Ingenier\'ia y Ciencias, Universidad Diego Portales, Av. Ej\'ercito Libertador 441, Santiago, Chile\\
$^{55}$ Instituto de Astrof\'{i}sica de Andaluc\'{i}a (CSIC), Glorieta de la Astronom\'{i}a, s/n, E-18008 Granada, Spain\\
$^{56}$ Departament de F\'isica, EEBE, Universitat Polit\`ecnica de Catalunya, c/Eduard Maristany 10, 08930 Barcelona, Spain\\
$^{57}$ Universit\'{e} Clermont-Auvergne, CNRS, LPCA, 63000 Clermont-Ferrand, France\\
$^{58}$ Department of Physics and Astronomy, Sejong University, 209 Neungdong-ro, Gwangjin-gu, Seoul 05006, Republic of Korea\\
$^{59}$ Abastumani Astrophysical Observatory, Tbilisi, GE-0179, Georgia\\
$^{60}$ Department of Physics, Kansas State University, 116 Cardwell Hall, Manhattan, KS 66506, USA\\
$^{61}$ Max Planck Institute for Extraterrestrial Physics, Gie\ss enbachstra\ss e 1, 85748 Garching, Germany\\
$^{62}$ Department of Physics, University of Michigan, 450 Church Street, Ann Arbor, MI 48109, USA\\
$^{63}$ University of Michigan, 500 S. State Street, Ann Arbor, MI 48109, USA\\
$^{64}$ Department of Physics, Southern Methodist University, 3215 Daniel Avenue, Dallas, TX 75275, USA\\
}

\end{document}